\documentclass{article}
\usepackage{jheppub} 
\usepackage{tikz}
\usetikzlibrary{decorations.pathmorphing}
\definecolor{darkmagenta}{rgb}{0.55, 0.0, 0.55}
\definecolor{darkblue}{rgb}{0.0, 0.0, 0.55}
\definecolor{darkred}{rgb}{0.7, 0.0, 0.3}

\def\Real{{\mathbb R}}

\def\1{1\hspace{-4pt}1}
\def\j1{\widetilde{1\hspace{-4pt}1}}
\usepackage{bbm}

\def\bec{\begin{center}}
\def\ec{\end{center}}

\let\oldsqrt\sqrt
\def\sqrt{\mathpalette\DHLhksqrt}
\def\DHLhksqrt#1#2{%
\setbox0=\hbox{$#1\oldsqrt{#2\,}$}\dimen0=\ht0
\advance\dimen0-0.2\ht0
\setbox2=\hbox{\vrule height\ht0 depth -\dimen0}%
{\box0\lower0.9pt\box2}}

\def\Rindler{{\mathfrak R}}
\def\cL{{\cal L}}

\def\cD{{\cal D}}

\def\cO{{\cal O}}

\def\cD{{\cal D}}
\def\cI{{\cal I}}

\def\cQ{{\cal Q}}

\def\cH{{\cal H}}

\def\cH{{\cal H}}
\def\cZ{{\cal Z}}
\def\q{q}

\def\d{{\rm d}}
\def\R{{\mathfrak{R}}}

\def\cO{{\cal O}}

\def\ed{\end{document}}

\def\bea{\begin{eqnarray}}
\def\eea{\end{eqnarray}}
\def\ba{\begin{array}}
\def\ea{\end{array}}

\def\tfrac#1#2{{\textstyle{{\scriptstyle #1}
\over {\scriptstyle #2}}}}

\definecolor{rougef}{rgb}{0.7,0,0}
\definecolor{vertf}{rgb}{0,0.6,0}
\definecolor{bleuf}{rgb}{0,0,0.9}

\usepackage[T1]{fontenc} 
\usepackage{graphicx,amssymb,amstext}
\usepackage{appendix}
\usepackage{float}
\usepackage{empheq}
\usepackage{mathtools}
\makeatletter
\allowdisplaybreaks

\usepackage[parfill]{parskip}
\newcommand{\be}{\begin{equation}}
\newcommand{\ee}{\end{equation}}

\newcommand{\vast}{\bBigg@{4}}
\newcommand{\Vast}{\bBigg@{5}}

\preprint{}

\title{Logarithmic Corrections, Entanglement Entropy, \\ and UV Cutoffs in de Sitter Spacetime}

\author[a]{Gabriel Arenas--Henriquez}
\author[b]{Felipe Diaz}
\author[c]{Per Sundell}

\affiliation[a]{Centre for Particle Theory, Department of Mathematical Sciences, Durham University, South Road, Durham, DH1 3LE, UK.}
\affiliation[b]{Departamento de Ciencias F\'isicas, Universidad Andres Bello, Sazi\'e 2212, Santiago de Chile.}
\affiliation[c]{Centro de Ciencias Exactas, Universidad del B\'io-B\'io, Avda. Andr\'es Bello 720, 3800708, Chill\'an, Chile.}

\emailAdd{gabriel.arenas-henriquez@durham.ac.uk} 
\emailAdd{f.diazmartinez@uandresbello.edu}
\emailAdd{per.anders.sundell@gmail.com} 

\abstract{ It has been argued that the entropy of de Sitter space corresponds to the entanglement between disconnected regions computable by switching on a replica parameter $q$ modeled by the quotient dS$/\mathbb{Z}_q$. 
Within this framework, we show that the centrally-extended asymptotic symmetry algebra near the cosmic horizon is a single copy of the Virasoro algebra. 
The resulting density of states matches the semi-classical result of Gibbons and Hawking up to an undetermined constant that is chosen to reproduce the entanglement entropy previously found in the literature. 
It follows that the logarithmic quantum corrections to the Cardy entropy reproduces the known one-loop result computed in the bulk in the presence of a cutoff. 
The resulting entanglement entropy follows the divergent area law, where the UV cutoff is now a function of the replica parameter.  
Thus, as the near-horizon CFT fixes the cutoff in units of the Planck scale, the model can be viewed as a probe into whether the defect Hilbert space has a finite dimension; indeed, the limit $q\to 0$, reproduces Banks' formula. 
We also study the quantum corrections of the effective description of the horizon entropy by means of Liouville field theory, where the large $q$ limit corresponds to a realization of dS$_3$/CFT$_2$ correspondence matching the logarithmic corrections to three-dimensional de Sitter space obtained by computing the one-loop contribution to the quantum gravity partition function in the round three-sphere.}

\begin{document}
\maketitle
\section{Introduction and summary}

Four-dimensional de Sitter spacetime, dS$_4$, is a vacuum solution to Einstein equations with a positive cosmological constant, with maximal isometry group, $SO(1,4)$, which acts as a positive vacuum pressure. 
This solution can be interpreted as a FLRW universe with positively curved spatial slices given by round three-spheres and exponential scale factor. 
Current observations show that our universe is described by a cosmological constant with positive small value, and is expanding with accelerating pace \cite{SupernovaCosmologyProject:1998vns, SupernovaSearchTeam:1998fmf}, such that at late times the universe may end in an exponential growth which is characterized by the dS$_4$ geometry. 
Early periods of the universe may have been dominated by an inflationary era after the big bang such that certain problems in cosmology can be understood by means of this inflationary process \cite{Linde:1981mu, Linde:2007fr}; thus, the understanding of de Sitter spacetime has a profound implication in the study of cosmology of our universe. 
One consequence of this exponential growth is that a static observer inside a dS$_4$ universe does not have access to the whole spacetime being limited by the existence of a cosmological horizon. Interestingly, this horizon is a Killing horizon which shares properties with the event horizon of a black hole.  For instance, it has been shown that there is an associated temperature \cite{Figari:1975km} analogue to Hawking radiation \cite{Hawking:1975vcx}, and therefore an entropy that was found by Gibbons and Hawking \cite{Gibbons:1977mu} which follows the area law \cite{PhysRevD.7.2333}. Nonetheless, in contrast with black hole thermodynamics, these are observer dependent quantities, and due to the lack of access to the asymptotic boundary, an static observer is limited  in constructing observables \cite{Witten:2001kn, Banks:2002wr}. 

The understanding of de Sitter spacetime poses a challenge to quantum gravity, and in particular to string and M theory, because of complications in constructing stable de Sitter solutions in supergravity \cite{Maldacena:2000mw, Kachru:2003aw}, which has  stimulated the idea that de Sitter spacetime may actually belong to  an untractable swampland of a string theory landscape \cite{Vafa:2005ui, Garg:2018reu}. 
As a matter of fact, the treatment of asymptotically de Sitter backgrounds in the context of M-theory has lead to the proposal that the dimension of the Hilbert space of a quantum theory of gravity is finite \cite{Banks:2000fe} given by the exponential of the Gibbons--Hawking entropy. This implies a problem for the quantization of gravity with Einstein--Hilbert action as a starting point \cite{Witten:2001kn}~. 

In \cite{Carlip_1999} has been shown, using covariant phase-space methods, that the constraint algebra of general relativity, upon taking a suitable set of boundary conditions, gives rise to a centrally extended, affine algebra.
Particularly, if the boundary consists of a Killing horizon, it is possible to relate this  algebra with one copy of the Virasoro algebra. 
Therefore, one can make use of conformal field theory (CFT) techniques in order to describe underlying degrees of freedom of the horizon. 
In the case of the black hole horizon, the Bekenstein--Hawking entropy matches the one of the horizon CFT \cite{Cardy:1986ie}. 
Remarkably, this result can be easily applied to the cosmological horizon of de Sitter space, in which the Cardy entropy associated to the horizon CFT renders the result obtained by Gibbons and Hawking.
Thus, by examining the asymptotic symmetry algebra of the horizon, one may obtain non-trivial information about the density of states of a Killing horizon, which thus extends the statistical mechanical approach to general relativity.
Moreover, computing Gaussian fluctuations, Carlip showed \cite{Carlip:2000nv} that the density of states acquires an extra contribution resulting in a logarithmic correction to the entropy of the horizon CFT. This reproduces the lowest order quantum corrections to black hole entropy that has been computed using other methods \cite{Kaul:2000kf, Das:2001ic}. 
In this approach, the asymptotic algebra is constructed by considering diffeomorphisms that preserve the asymptotic structure and have well defined asymptotic charges associated to them.
In this construction, the central charge appears with a free parameter that can be fixed in order to recover the area law.
This mechanism has been used \cite{Mahapatra:2016iok} in the context of the AdS/CFT correspondence such that the holographic Renyi entropy acquires an extra contribution that imply a logarithmic divergence in the entanglement limit. 

Recently, it has been argued \cite{Kabat:2002hj,Narayan:2015vda, Nomura:2017fyh,Narayan:2017xca, Dong:2018cuv, Arias:entanglement, Arias:liouv, Narayan:2019pjl,Fernandes:2019ige, Narayan:2020nsc, Geng:2020kxh,  Manu:2020tty,Goswami:2021ksw, Susskind:2021esx, Shaghoulian:2021cef, Shaghoulian:2022fop} by using different approaches, that de Sitter space corresponds to a maximally entangled state incorporating pairings between disconnected regions of spacetime, leading to a density matrix with an associated Renyi entropy that is constant.
Therefore, one can interpret the Gibbons--Hawking entropy as the entanglement entropy between the entangled regions.
In \cite{Arias:entanglement, Arias:liouv}, the entanglement between disconnected Rindler observers has been computed by using a single coordinate patch that contains two causally disconnected Rindler observers, and considering the reduced density matrix of one of them.
This was achieved by letting  $\mathbb{Z}_q$ act on the doubled Rindler patch so as to produce an orbifold geometry with two antipodal conical defects\footnote{In principle, one may assign the two pole separate $q_{N}$ and $q_{S}$ parameters leading to a Thurston's spindle.
In our work, we take $q_{N}=q_{S}=q$ such that the resulting geometry can be described by means of a single quotient dS$/\mathbb{Z}_q$~.}.
The parameter $q$ can be identified as the replica parameter and the $q\to1$ limit, referred to as the tensionless limit, leads back to the smooth de Sitter geometry. 
This limit is nothing else than the entanglement entropy limit of the Renyi entropy, which is bigger than the Gibbons--Hawking entropy due to the presence of both patches considered \cite{Arias:entanglement}. Moreover, following \cite{Solodukhin:1998tc}, a consistent truncation of this construction onto Liouville field theory gives a $q$-dependent central charge that vanishes in the tensionless limit, thus having zero Cardy entropy. The associated Renyi entropy of the system equals the Gibbons--Hawking entropy in agreement with previous results. 

In this paper, we combine Carlip's conformal description of the cosmological horizon by imposing maximal entanglement requiring $q$-independent Renyi entropy which leaves one free parameter that can be used to match the entanglement entropy with the Gibbons--Hawking entropy at the semi-classical level. 
To this end, we use of the Barnich--Brandt formalism \cite{Barnich:2001jy, Barnich:2006av} to compute the asymptotic charges using the Carlip--Silva asymptotic generators \cite{Carlip:1999cy, Silva:2002jq}. This preserves the asymptotic structure and produces finite charges whose algebra yields the same non-trivial central charge as that found by Carlip \cite{Carlip:1999cy}.
Assuming furthermore, the validity of Cardy entropy formula, the Gibbons--Hawking entropy is recovered upon matching the free parameter.
We then consider the quantum corrections to the Cardy entropy found in \cite{Carlip:2000nv}.
It follows that the Renyi entropy is no longer constant and diverges in the tensionless limit as a direct consequence of the fact that the central charge vanishes in this particular point. 
We also investigate other limits of the corrected Renyi entropy: the min-entropy limit, which gives the larges eigenvalue of the reduced density matrix; the max-entropy which gives the dimension of the Hilbert space, and the von Neumann entropy, that is, the entanglement entropy. 
We show that the latter now acquires a UV divergence in agreement with the entanglement area law in quantum field theory (QFT) due to the short-range interaction near the boundaries of the entanglement region \cite{Srednicki:1993im}. This divergence arises in the tensionless limit (that as aforementioned, corresponds to the entanglement entropy limit), allowing us to introduce a UV cutoff giving a physical interpretation for the orbifold parameter. 
The IR finite term corresponds to the Gibbons--Hawking entropy, and we show that the proposed connection between the dimension of the Hilbert space and the entropy holds beyond the semiclassical order by noticing that the cutoff goes to infinity in the max-entropy limit.

In \cite{Arias:liouv}, it was shown that the large $q$ limit resembles the three-dimensional global dS geometry with the defects corresponding to the conformal future- and past-infinities $\cI^\pm$. 
Therefore, the Liouville central charge resembles the one found in \cite{Strominger:2001pn} of $\cI^\pm$ by considering an holographic CFT description of dS in three dimensions. 
The corresponding Cardy entropy renders the Gibbons--Hawking entropy. 
We also consider the logarithmic corrections to this scheme, showing that in the $q\to\infty$ limit, the resulting min-entropy corresponds to the three-dimensional de Sitter entropy with the quantum corrections recently found in \cite{Anninos:2020hfj, Coleman:2021nor}.

\section{Rindler horizons and defects}

\subsection{Undeformed de Sitter geometry} 

\paragraph{Global geometry.}

Global de Sitter spacetime of signature $(1,d-1)$ and radius $\ell$, here denoted by dS$_d$, is the solution to Einstein's field equations in $d$ dimensions with cosmological constant $\Lambda=(d-1)(d-2)/(2\ell^2)$ given by the induced metric $ds^2$ on the hyperboloid 
\be\label{hyper}
Y_A Y^A=\ell^2~,\qquad A=0,1,\dots,d~,
\ee
in $(d+1)$-dimensional Minkowski spacetime, viz.
\begin{align}
    ds^2 = -dY_0^2 + \sum_{i=1}^d dY_i^2 ~.
\end{align}
The global geometry is maximally symmetric with isometry group $O(1,d-1)$, and can be viewed as a foliation along a global time with the topology of $\mathbb{R}$ with space-like leaves given by round $(d-1)$-spheres with two conformal boundaries at $Y_0\to \pm\infty$, denoted by ${\cal{I}}^\pm_d$, respectively.
The intrinsic line-element can be coordinatized as 
\begin{align}
    ds^2 = -dT^2 + \ell^2\cosh^2(T/\ell)d\Omega_{d-1}^2\ ,
\end{align}
using a global time coordinate $T\in ]-\infty,+\infty[$, and the metric $d\Omega_{d-1}^2$ of the round, unit $(d-1)$-sphere.

\paragraph{Causal patches.}
The exponential inflation prevents a static observer, $O$, say, often referred to as a Rindler observer, from communicating with the entire geometry, since upon emitting light rays, these must return to $O$ prior to having reached half-way across the spatial slice.
The envelop of all such rays forms a causally connected region, $\Rindler_O\subset {\rm dS}$, referred to as a Rindler patch.
Each such patch has $S^2$-topology and a metrically bifurcated boundary
\be \partial \Rindler_O={\cal H}_- \cup ( {\cal H}_\infty\times \mathbb{R}_-)\cup {\cal H}_\infty\cup  ({\cal H}_\infty\times \mathbb{R}_+)\cup {\cal H}_+\ ,\qquad  {\cal H}_\pm\cong {\cal H}_\infty \cong S^{d-2}\ ,\label{2.4}\ee
where ${\cal H}_\infty$ are the cosmological horizons, with past and future limits ${\cal H}_-$ and ${\cal H}_+$, respectively, residing as defects inside ${\cal I}^\pm$.
The interior of $\Rindler_O$ as well as the branches of its bifurcated boundary are stabilized by
\be SO(1,d)_{\Rindler}\cong SO(1,d-1)\ ,\ee
which together with the discrete time-reversal operation makes up the maximal isometry group $O(1,d-1)$ of a Rindler patch.

\paragraph{Euclidean patch.}

Treating de Sitter path integral in the Euclidean regime, the required Wick rotation maps each Rindler patch into a round $S^d$, and the Euclidean field theory has a Hawking temperature \cite{Figari:1975km} 
\begin{align}
    T_{\rm H} = \frac{\hbar}{2\pi\ell}~,
\end{align}
with conjugated Gibbons--Hawking entropy \cite{Gibbons:1977mu}
\begin{align}
    S_{\rm GH} = \frac{A_\cH}{4\hbar G_d}~,
\end{align}
where $A_\cH$ is the area of the cosmological horizon, and $G_d$ is Newton's constant in $d$ dimensions.

\paragraph{Double Rindler patch.}

From here on, we will work in four space-time dimensions, where $A_\cH = 4\pi\ell^2$~. 
In order to exhibit entanglement, it is natural to introduce coordinates adapted to unions $\Rindler_S\cup \Rindler_N$ of pair of causally disconnected Rindler patches, which can be achieved by the embedding \cite{Arias:entanglement}
\begin{align}
\label{embedding1}
Y_0 = &\sqrt{\ell^2 -r^2}\,\cos\theta\,\sinh(t/\ell)~,\quad
Y_1 =\sqrt{\ell^2 - r^2}\,\cos\theta\,\cosh(t/\ell)~,  \\ \nonumber
Y_2 &= r\,\cos\theta ~,\quad
Y_3 =\ell \sin\theta \cos\phi~,\quad
Y_4 = \ell \sin\theta \sin\phi ~,
\end{align}
where 
\begin{equation}
-\infty<t<\infty~,\quad 
-\ell<r<\ell~,\quad 
0\leq \theta\leq\pi~,\quad  
0\leq\phi<2\pi~,
\end{equation}
with $N$ and $S$ tracing out the worldlines $(r,\theta)=(0,0)$ and $(r,\theta)=(0,\pi)$, respectively.
The resulting intrinsic line element, viz.
\begin{align}\label{dS}
    ds^2\rvert_{\Rindler_S\cup \Rindler_N} = \ell^2(d\theta^2 + \sin^2\theta d\phi^2) + \cos^2\theta\left(-f(r)dt^2 +\frac{dr^2}{ f(r)}\right)~,\qquad f(r) = 1-\frac{r^2}{\ell^2}\ ,
\end{align}
is a fibration over $S^2$ with fibers above the equator given by points, and fibers above the strictly northern and southern hemispheres given by radially extended, two-dimensional Rindler patches, denoted by $\Rindler_2$, which are indeed contractible to points. This is illustrated in Figure~\hyperlink{Fig:1}1~.
\begin{center}
\begin{tikzpicture}[scale=0.7]
\fill[fill=yellow!10] (0,0)--(2,2)--(0,4);
\fill[fill=blue!10] (4,0)--(2,2)--(4,4);
\draw [thick] (0,4) --node[above] {$\mathbf{\mathcal I^+}$}(4,4);
\draw [thick] (0,0) --node[below] {$\mathbf{\mathcal I^-}$}(4,0);                        
\draw [thick] (4,0)--(4,4); 
\draw [thick] (0,0) --(0,4);
\node at (1,2) {$\mathfrak R_N$};
\node at (3,2) {$\mathfrak R_S$};
\draw [thick, dashed] (0,0) --(4,4); 
\draw [thick, dashed] (0,4)--(4,0);
\node[text width=13cm, text justified] at (2,-2){
\small{\hypertarget{Fig:1}\bf Fig.~1}: 
\sffamily{
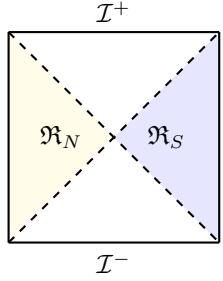
Penrose diagram of de Sitter spacetime with two Rindler observers $\Rindler_N$ and $\Rindler_S$. \\ $\cal{I}^\pm$ corresponds to the past and future conformal infinities of the global geometry.
}};
\end{tikzpicture}
\end{center}

We denote the northern and southern fibers by $\Rindler_2\pm$, respectively, such that $N$'s worldline is located at $r=0$ in the $\Rindler^+_2$ above $\theta =0$, idem $S$ and $\theta=\pi$.
The stabilizers in $SO(1,4)$ of each embedded $\Rindler_2$-fiber is given by
\be SO(1,4)_{\Rindler_2^\pm}=SL(2,\mathbb{R})\times U(1)\ ,\ee
where the $SL(2,\mathbb{R})$-factor contains a $U(1)$ subgroup generated by the time-translation vector field $\vec\partial_t$~, and the $U(1)$-factor is generated by the rotational vector field $\vec \partial_\phi$\footnote{The corresponding isometry group matches the one of the near horizon geometry of an extremal Kerr black hole, such that the asymptotic symmetry group generators of \cite{Guica:2008mu} can be used in order to reproduce the Gibbons--Hawking entropy \cite{Diaz:2019khq}.}.

\subsection{Orbifolding the horizon}
We follow \cite{Arias:entanglement, Arias:liouv}, by considering the orbifold
\be dS_{q} := \widehat{(\Rindler_N\cup \Rindler_S)}_q/\mathbb{Z}_q\ ,\qquad f=f\sim g\sim \mathbb{Z}_{q}\subset SO(1,3)\ ,\ee
with $q$-fold, branched cover $\widehat{(\Rindler_N\cup \Rindler_S)}_q$ with a natural $\mathbb{Z}_q$-action via a Killing vector field $\vec J$ of length $g(\vec J,\vec J)=1$; as $\vec J$ generates a $U(1)$ subgroup of $SO(1,3)$, the orbifold parameter can be extended from the positive integers to $q\in \mathbb{R}_+$.
The orbifold fixed points form a pair $(\Sigma_S,\Sigma_N)$ of antipodal, sub-manifolds of co-dimension two,
referred to as defects, with induced $\Rindler_2$ metrics.
The defects generate $\delta^2$-function singularities in the Riemann curvature two-form \cite{Fursaev:1995ef} proportional to the conical deficit $2\pi\left(1-1/q\right)$.
A well-posed variational principle requires to supplement the Einstein--Hilbert action with extra terms localized at the defects that provide singular stress tensors cancelling the singularities in the bulk Einstein tensor, such that the field equations hold everywhere, even at the defects \cite{Gregory:1988xc}. To treat maximally symmetric defects, it suffices to dress the defects by Nambu--Goto actions with a $q$-dependent tension ${\cal T}_{q_\xi}$, resulting in a total action
\be \label{Iq} 
I_{q}=I_{\rm EH}+I_{\rm def}\ ,\qquad I_{\rm def} = \sum_{\xi = {\rm N,S}}I_{\xi}\ ,\ee\be I_{\xi}=-{\cal T}_{q_{\xi}} \int_{\Sigma_\xi} d^2 \sigma\sqrt{-{\rm det}(g|_{\Sigma_\xi})}~, \qquad {\cal T}_{q_\xi} = \frac{1}{4G_4}\left(1-\frac{1}{q_\xi}\right)\ ,\qquad q_{\rm N} = q_{\rm S} = q\ .
\ee
Thus, the smooth de Sitter geometry is recovered in the $q\to 1$ limit, referred as the tensionless limit. The dressed defects can thus be interpreted \cite{Arias:liouv} as Rindler observers with mass proportional to $1-q$, stretched out into strings (in the direction determined by $\vec J$) back-reacting to the surrounding geometry.
Choosing polar coordinates on $S^2$ such that $\vec J=\vec\partial_\phi$, the orbifolded line element 
\begin{align}\label{extended}
    ds^2_q = \cos^2\theta\left(-f(r)dt^2 + \frac{dr^2}{f(r)}\right) +\ell^2\left(d\theta^2 + \frac{\sin^2\theta}{q^2} d\phi^2\right)~,
\end{align}
where the ranges of all coordinates are as before, which thus cover the entire $q$-deformed, double Rindler patch, with Rindler defects located at $\theta=0$ and $\theta=\pi$, respectively; it follows that
\begin{align}
    ds^2_q\big\rvert_{\Sigma_{N,S}} = -f(r)dt^2 + \frac{dr^2}{f(r)}~. \label{h}
\end{align}
The $q$-deformed, double Rindler wedge has cosmological horizon given by a bifurcated Killing horizon $S^2$-topology located at $r=\pm\ell$, where the Killing vector field $\vec \chi:=\vec\partial_t$ passes from being time-like to light-like, with $q$-independent surface gravity, viz.
\be\label{chi}
\chi^2:= ds^2_q(\vec\chi,\vec \chi) = \frac{\cos^2\theta}{\ell^2}\left(r^2-\ell^2\right)~,\qquad \kappa^2 := -\lim_{\chi^2\to 0} \frac{\nabla_{\mu} \chi^2 \nabla^\mu \chi^2}{4\chi^2} = \frac{1}{\ell^2}\ .\ee
The orbifolding procedure of the horizon is illustrated in Figure~\hyperlink{Fig:2}2

\vspace{0.8cm}
\begin{center}
\begin{tikzpicture}[scale=0.9]
\draw [thick] (-4,0) circle (2cm);
\node [darkred] at (-4,2) {\textbullet};
\node [darkred] at (-4,-2) {\textbullet};
\node at (-6,2) {$\mathcal H\cong S^2$};
\draw[thick, dashed] (-2,0) 
arc[start angle=0,end angle=180, 
x radius=2, y radius=0.5];
\draw[thick] (-2,0) 
arc[start angle=0,end angle=-180, 
x radius=2, y radius=0.5];
\node at (-0.5,0.4) {$S^2/\mathbb Z_q$};
\draw [thick, ->] (-1.3,0) -- (0.3,0);
\fill[fill=yellow!10] (1,2.3)--(0,1.7)--(5,1.7)--(6,2.3);
\fill[fill=blue!10] (1,-1.7)--(0,-2.3)--(5,-2.3)--(6,-1.7);
\draw [thick] (1,2.3) --(6,2.3);
\draw [thick] (0,1.7) --(5,1.7);
\draw [thick] (0,1.7) --(1,2.3);
\draw [thick] (6,2.3) --(5,1.7);
\coordinate (N) at (3,2);\coordinate (S) at (3,-2);
\node [darkred] at (3,2) {\textbullet};
\node [darkred] at (3,-2) {\textbullet};
\draw[thick] (N)to[out=-20,in=20](S);
\draw[thick] (N)to[out=-150,in=150](S);
\draw[thick, dashed] (4.1,0) 
arc[start angle=0,end angle=180, 
x radius=1.05, y radius=0.3];
\draw[thick] (4.1,0) 
arc[start angle=0,end angle=-180, 
x radius=1.05, y radius=0.3];
\node at (6,1.7) {$\Sigma_N$};
\node at (0.1,-1.6) {$\Sigma_S$};
\draw [thick] (1,-1.7) --(6,-1.7);
\draw [thick] (0,-2.3) --(5,-2.3);
\draw [thick] (0,-2.3) --(1,-1.7);
\draw [thick] (6,-1.7) --(5,-2.3);
\draw [thick, ->] (3.05,0) -- (4.05,0);
\node at (5.1,0) {$\ell_q=\ell/q $};
\node[text width=13cm, text justified] at (0,-4) 
{\small {\hypertarget{Fig:2}\bf Fig.~2}: 
\sffamily{
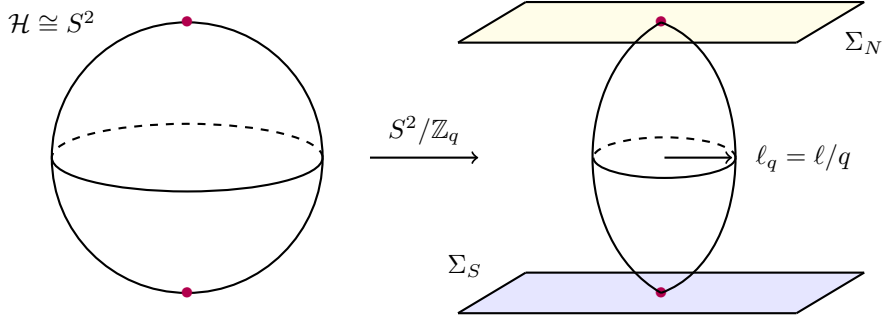
The spherical geometry of the cosmological horizon $\cH$ after the orbifolding deforms to that of the spindle $S^2/\mathbb{Z}_q$ with radius $\ell_q = \ell/q$. The set of fixed points $\Sigma_N$ and $\Sigma_S$ correspond to the defects located at $\theta = 0, \pi$ respectively, both with induced metric \eqref{h}. }
};
\end{tikzpicture}
\end{center}

\subsection{Semi-classical entropy computation }\label{Sec:2.4}

Using the formalism of \autoref{App:Entropies}, we assign to each Rindler observer a $q$-independent Hamiltonian $H_\Rindler$, and a reduced density matrix $\rho_\Rindler:=e^{-H_\Rindler}$.
Introducing a modular temperature 
\be T_q:=q^{-1}\ ,\ee
treated as the conjugate variable of $H_\Rindler$, the resulting thermal partition function
\begin{align}
\cZ_\Rindler := {\rm tr}_{H_\Rindler}\rho_\Rindler^q = {\rm tr}_{H_\Rindler}\,\exp\left\{-q H_\Rindler\right\}\ ,
\end{align}
computed on the $q$-fold, branched cover $\widehat S^4_q$ of the Euclideanized orbifold
\begin{align}
    S^4_q:= \widehat S^4_q/\mathbb{Z}_q\ ,
\end{align}
can be re-written using the Cardy--Calabrese formula \cite{Calabrese:2004eu}
\begin{align}
\cZ_\R =  \frac{\cZ[\widehat S_q^4]}{\cZ[S^4]^q}=\left(\frac{\cZ[ S_q^4]}{\cZ[S^4]}\right)^q\ .
\end{align}
In the semi-classical approximation,
\begin{align}
   \cZ[S_q^4]\approx \exp\left\{-\frac{1}\hbar I^E[S^4_q]\right\}\ ,\qquad  \cZ[S^4] \approx \exp\left\{-\frac1\hbar I^E[S^4]\right\}~,
\end{align}
where the total Euclidean action 
\begin{align}
    I^E_q[S^4_q] = -\frac{q}{16\pi G_4}\int_{S^4} d^4x\sqrt{{\rm det}(g_q)}\left(R-2\Lambda\right) + I^E_{\rm def} = \left(1-\frac2q\right)\frac{\pi\ell^2}{G_4}~.
\end{align}
It follows that each $q$-deformed Rindler observer has the modular free energy
\begin{align}\label{freeq}
    F_q := -\frac{1}{q}\log\cZ_\Rindler \approx \frac{1}\hbar (I^E_\Rindler [S^4/\mathbb{Z}_q] - I^E[S^4] )\nonumber  = 2\left(1-\frac{1}{q}\right)\frac{\pi\ell^2}{\hbar G_4}~,
\end{align}
which is the quantity to be identified with the semi-classical entropy obtained by computing the thermal partition function in reduced phase-spaces attached to various regions of the Rindler patch.
From the exact relation between Renyi entropy and free energy, 
\be\label{Baez}
S_q = \left(1-\frac1q\right)^{-1}F_q~,
\ee
it follows that 
\begin{align}
    S_q \approx \frac{2\pi\ell^2}{\hbar G_4} = 2S_{\rm GH}~,\qquad \forall q\geqslant 1\ ,\label{2.26}
\end{align}
whose $q$-independence can be interpreted as that the original pair of Rindler observers formed a  maximally entangled state in agreement with \cite{Kabat:2002hj,Nomura:2017fyh,Dong:2018cuv, Arias:entanglement, Arias:liouv, Narayan:2019pjl, Susskind:2021esx, Shaghoulian:2021cef, Shaghoulian:2022fop}, supporting the interpretation of that the Gibbons--Hawking entropy is due to entanglement between causally disconnected Rindler patches\footnote{In \cite{Arias:liouv}, the modular free energy was instead computed by means of a consistent truncation of the total action to a Liouville theory on the defect, which indeed yields a constant Renyi entropy in the semiclassical regime as well.}.

\section{Near-Horizon algebra and Cardy entropy}\label{Sec:CentralCharge}

\subsection{Asymptotic charges}

Carlip has shown \cite{Carlip:1999db,Carlip:1999cy, Carlip_1999} that general relativity admits classes of asymptotic boundary conditions near Killing horizons forming modules of centrally extended, asymptotic Virasoro charges. Under the assumption of modular invariance, the asymptotic charges induce an entropy obeying the area law.

In what follows, we shall assign the $q$-deformed Rindler patches classes of boundary conditions at their cosmological horizons inducing Virasoro modules with finite central charges for $q>1$ fixed up to one free parameter by assumption of maximal entanglement.
These considerations are facilitated by the space-time covariant Barnich--Brandt--Compere (BBC) formalism \cite{Barnich:2001jy, Barnich:2006av, Barnich:2007bf, Compere:2007az}, which maps algebras of asymptotic Killing vector fields (AKVs) $\vec\zeta$ stabilizing spaces of metric fluctuations $h$ in the near-cosmic horizon region to charges
\begin{align}\label{Qalg}
    \cQ_{\vec\zeta}[h;g] = \frac{1}{8\pi G_4}\oint\displaylimits_{\cH}  k_{\vec\zeta} [h,g]~,
\end{align}
where $\cH$ denotes the cosmological horizon, and\footnote{Tensorial indices are raised and lowered using the background metric $g$.} 
\begin{align}
    k_{\vec\zeta}[h,g] = -\frac14dx^\alpha \wedge dx^\beta\epsilon_{\alpha\beta\mu\nu}\Big[&\zeta^\nu\nabla^\mu h^\rho{}_\rho - \zeta^\nu\nabla_\sigma h^{\sigma\mu} + \zeta_\sigma\nabla^\nu h^{\mu\sigma}  \\&+ \frac12 h^\rho{}_\rho\nabla^\nu \zeta^\mu - h^{\nu\sigma}\nabla_\sigma \zeta^\mu \nonumber  +\frac12 h^{\sigma\nu}\left(\nabla^\mu \zeta_\sigma + \nabla_\sigma \zeta^\mu \right)\Big]~,
\end{align}
is a background-covariant two-form, obeying the centrally extended, Dirac-bracket algebra
\begin{align}\label{DiracExt}
    \left\{\cQ_{\vec\zeta},\cQ_{\vec\zeta'}\right\} =  \cQ_{[\vec\zeta',\vec\zeta]} +\frac{1}{8\pi  G_4}\int\displaylimits_\cH  k_{\vec\zeta} [\cL_{\vec\zeta'}g,g]~,
\end{align}
where $\cL_{\vec\zeta}$ denotes the Lie derivative along $\vec\zeta$.
The two-form is a potential of the Noether three-form \cite{Julia:1998ys}, which is closed on the entire Rindler patch, including the defect, as a consequence of the variational principle, which means that the central charge accounts for those asymptotic degrees of freedom that extend smoothly into the entire Rindler patch (including the defect).

A family of vectors satisfying the Diff$(S^1)$ algebra (or de Witt algebra) 
\begin{align}\label{Witt}
    i[\vec{\zeta}_m,\vec{\zeta}_n] = (m-n)\vec{\zeta}_{m+n}~,
\end{align}
and that preserve the asymptotic symmetries of a Killing horizon, has been found in \cite{Carlip:1999cy, Silva:2002jq} and have the form of
\begin{align}\label{akvs}
    \zeta^\mu_m = T_m {\chi}^\mu+R_m \rho^\mu~.
\end{align}

Here $\chi^\mu$ is the null vector defined in \eqref{chi}, while $R_m$ and $T_m$ are coordinate-dependent functions obtained by requiring that the structure of the horizon is preserved \cite{Carlip_1999}, i.e., $\chi^{\mu}$ remains null at $r=\ell$, and 
\begin{align}
\rho_\mu = -\frac{1}{2\kappa}\nabla_\mu \chi^2~,
\end{align}
is a vector normal to the horizon with vanishing norm at the horizon.
To obtain the Killing vectors $\vec\zeta$~, is enough to assume that, near the horizon, they correspond to conformal Killing vectors \cite{Solodukhin:1998tc, Silva:2002jq}, viz.
\begin{align}
    \nabla_\mu\chi_\nu + \nabla_\nu\chi_\mu = \cO(\chi^2)g_{\mu\nu}+\cO(\chi^4)~,
\end{align}
such that a set of diffeomorphisms $\vec{\zeta}_m$ of \eqref{dS} must satisfy
\begin{align}
    \delta_\zeta \chi^2 ={}& \cO(\chi^4)~, \\ \delta_\zeta \rho_\mu ={}& \cO(\chi^2)\rho_\mu~,
\end{align}
that for \eqref{akvs} implies
\begin{align}
    \cL_\chi T_n + \kappa R_n = \cO(\chi^2)~,\qquad \rho^\mu\nabla_\mu T_n = \cO(\chi^2)T_n~.
\end{align}
These asymptotic generators, indeed, satisfy a stronger condition imposed by Carlip \cite{Carlip:1999cy}
\begin{align}\label{Carlip}
    \lim_{\chi^2\to0}\frac{\chi^\mu\chi_\nu}{\chi^2}\nabla_\mu\zeta^\nu_m =\lim_{\chi^2\to0} \chi^\mu\nabla_\mu\left(\frac{\chi_\nu\zeta^\nu_m}{\chi^2}\right)-\kappa\frac{\rho_\mu\zeta_m^\mu}{\chi^2} = 0~,
\end{align}
guaranteeing that $\chi^{\mu}$ is a Killing vector in the horizon neighborhood. 
Therefore, the condition \eqref{Carlip} can be seen as the horizon analogs to the fall-off conditions that one uses at infinity to obtain the asymptotic symmetry algebra. 
Particularly, the previous condition is satisfied for \eqref{akvs} with
\begin{align}
T_m = -\frac{\ell}{\alpha}\exp\left\{im(\phi-\alpha t/\ell)\right\}~,\qquad R_m = \frac{\alpha}{l\kappa}imT_m~,
\end{align}
where $\alpha$ is an undetermined real number, satisfying the de Witt algebra \eqref{Witt} and preserving the asymptotic structure. The $\alpha$ parameter has been chosen in order to reproduce the horizon entropy in \cite{Carlip:1998qw, Carlip:1999cy, Carlip:1999db, Carlip:2000nv, Silva:2002jq}. In the following subsection, we will rescale $\alpha$ by requiring that the resulting Renyi entropy describe a maximally mixed entangled state as obtained in \cite{Dong:2018cuv, Arias:entanglement}, such that $\alpha$ depends on the orbifold parameter $q$, giving a central charge that resembles the one found in \cite{Arias:liouv}.

\subsection{Asymptotic Virasoro symmetries}

The resulting symmetry group generated by the AKVs \eqref{akvs} is given by a single copy of \eqref{DiracExt} with central extension
\begin{align}
    \frac{1}{8\pi G_4}\int\displaylimits_{\cH} k_{\vec\zeta_m}[\cL_{\vec\zeta_n}g,g] = -i\left(\frac{\ell^2}{4q\alpha G_4}\right)(\alpha^2 m^3+2m)\delta_{m+n,0}~.
\end{align}
Thus, promoting the Dirac bracket to a quantum commutator, viz. $\{\cdot,\cdot\}\to\tfrac{1}{i\hbar}[\cdot,\cdot]$~, and defining quantum operators
\begin{align}
     L_m := \cQ_m + \frac{3\ell^2}{8 G_4}\frac{\alpha}{q}\delta_{m,0}~,
\end{align}
yields a chiral Virasoro algebra
\begin{align}
    [L_m, L_n] = (m-n)L_{m+n} + \frac{c_q}{12}m(m^2-1)\delta_{m+n,0}~,
\end{align}
with central charge 
\begin{align}\label{c}
    c_q = 12i \lim_{r\to\ell} \cQ_{\vec\zeta_m}[\cL_{\vec\zeta_{-m}}g,g]\big\rvert_{m^3} = \frac{3\ell^2}{\hbar G_4}\frac{\alpha}{q}~.
\end{align}

\subsection{Maximal entanglement condition}

Assuming that the asymptotic charges give rise to a modular invariant thermal quantum theory, the semi-classical approximation of the corresponding Virasoro partition yields the Cardy entropy \cite{Cardy:1986ie}
\begin{align}\label{Cardy}
    S_{\rm C} \approx S_{\rm C}^{(0)}:= 2\pi\sqrt{\frac{c_q \Delta}{6}} = \frac{\pi^2}{3}c_q T\ ,
\end{align}
where $T$, which is the temperature of the chiral modes near the horizon, has been eliminated by means of 
\begin{align}
    \frac{\partial S_{\rm C}}{\partial \Delta} = \frac{1}{T}~\quad \Rightarrow\quad \Delta = \tfrac{\pi^2}{6}c_q T^2\ . 
\end{align}
Viewing the orbifolding procedure as a computation in a quantum theory using replicas, the Cardy entropy \eqref{Cardy} is identified as the modular free energy $F_q$ of the degrees of freedom in the $q$-deformed near-horizon region.
Let us redefine $\alpha = 2\gamma(q-1)$~, where $\gamma>0$ is a finite undetermined constant, by requiring that the original pair of Rindler observers formed a maximally mixed entangled state, that is, that the corresponding Renyi entropy $S_q$, which is given by Baez' formula \eqref{Baez}, is $q$-independent.  Therefore, the central charge, Cardy entropy, and Renyi entropy are given by
\begin{align}\label{B}
     c_q =  \frac{6 \ell^2}{\hbar G_4}\gamma\left(1-\frac{1}{q}\right)\ ,\qquad      S^{(0)}_{\rm C} = \left(1-\frac{1}{q}\right)\gamma S_{\rm GH}~,\qquad   S^{(0)}_q = \gamma S_{\rm GH}~,\ \end{align}
respectively. When $\gamma = 1$, the above result agrees with previous computations \cite{Kabat:2002hj,Nomura:2017fyh, Dong:2018cuv, Arias:entanglement,Shaghoulian:2022fop}. Therefore, the cosmological horizon of the deformed geometry reduces to that of the smooth dS geometry, whose thermal properties can thus be interpreted as being due to the entanglement between the causally disconnected Rindler observers encoded into the two-dimensional thermal CFT with modes having temperature \eqref{T} and central charge \eqref{c}. 

\section{Logarithmic corrections and UV divergences}\label{Sec:Log}
\subsection{$q$-dependent bulk cut-offs}

Working perturbatively inside the bulk, on a single Wick-rotated Rindler patch, $\Rindler$, the free energy can be expanded as 
\be \cZ_\Rindler \approx \exp\left\{-\tfrac{1}{\hbar}I^E_q -\frac12\log{\rm det}\nabla\right\}\ ,\label{4.2}\ee
including the semi-classical  Gibbons--Hawking entropy and its one-loop correction.
By our hypothesis, the thermal description of the near-horizon entropy is captured by a chiral CFT with Bunch--Davies temperature and central charge $c_q$ given in Eq. \eqref{B}.
We also recall that in a thermal replica theory with normalized inverse temperature $q$, the Renyi entropy $S_q$ is related via \eqref{Baez} to the moduluar, free energy $F_q$ in its turn given by 
\be
F_q = -\frac1q\log\cZ_\Rindler = S_{\rm C}~;
\ee
for further details, see \autoref{App:Entropies}.
Including Carlip's correction to the Cardy entropy \cite{Carlip:2000nv}, the density of states for a two-dimensional, thermal, chiral CFT can be written as
\begin{align}
    \rho(\Delta)\approx \left(\frac{c}{96\Delta^3}\right)^\frac14 \exp\left\{2\pi\sqrt{\frac{c\Delta}{6}}\right\}\quad \leftrightarrow\quad 
    \rho(T) \approx \frac{c}{144}(S_{\rm C}^{(0)})^{-\frac32}\exp\left\{S_{\rm C}^{(0)}\right\}~.
\end{align}
Using the central charge $c_q$ for the asymptotic charges in the near-horizon region, and the corresponding semi-classical Cardy entropy $S_{\rm C}^{(0)}$, both given in \eqref{B}, the logarithmically corrected Cardy entropy
\begin{align}
    S_{{\rm C}} \approx S^{(0)}_{\rm C} - \frac{3}{2}\log S^{(0)}_{\rm C} + \log\left(\frac{c}{144}\right)  = \left(1-\frac1q\right) \gamma S_{\rm GH} - \frac32\log\left[\left(1-\frac1q\right)^{\frac13} \gamma S_{\rm GH}\right]~,
\end{align}
modulo constants, which thus contains a divergent quantum correction in the tensionless limit.
Following \autoref{Sec:2.4}, we can obtain the modular free energy by inverting \eqref{Baez}, which yields the Renyi entropy
\begin{align}\label{Renyi}
   S_{q} ={}& \gamma S_{\rm GH} - \frac{3}{2}\left(\frac{q}{q-1}\right)\log\left[\left(1-\frac{1}{q}\right)^{\frac13} \gamma S_{\rm GH} \right]~,
\end{align}
with the following limits
\begin{align}
    S_{\rm E} ={}& \lim_{q\to 1}S_q = \gamma S_{\rm GH} -\frac{3}{2}\lim_{q\to1} \frac{q}{q-1}\log\left[\left(1-\frac{1}{q}\right)^\frac13 \gamma S_{\rm GH}\right]~,\label{Entanglement} \\[0.9ex] S_\infty ={}& \lim_{q \to \infty}S_q = \gamma S_{\rm GH} - \frac{3}{2}\log \gamma S_{\rm GH}~, \label{min}\\[0.9ex]  S_{0} ={}& \lim_{q\to 0} S_q = \gamma S_{\rm GH} \label{zero}~.
\end{align}

As we can see from \eqref{Entanglement}, the entanglement entropy acquires an extra divergent term coming from the 1-loop determinant and a finite term corresponding to the Gibbons--Hawking entropy as previously computed in different contexts \cite{Kabat:2002hj,Nomura:2017fyh, Dong:2018cuv, Arias:entanglement, Arias:liouv, Narayan:2019pjl, Susskind:2021esx, Shaghoulian:2021cef, Shaghoulian:2022fop}.
The extra logarithmic divergence can be rewritten as
\begin{align}
    S_{\rm E} = \gamma\frac{A_\cH}{\delta_q^2} + \gamma\frac{A_\cH}{4\hbar G_4}~,
\end{align}
where $\delta_q$ reads as 
\begin{align}
   \delta_q^2 = \frac{2\gamma A_\cH}{3}\left(\frac{1-q}{q}\right) \log\left[\left(1-\frac1q\right)^\frac13 \frac{\gamma A_\cH}{4\hbar G_4}\right]^{-1}\quad  \leftrightarrow\quad \left(1 - \frac{1}{q}\right) = \frac{\delta^2_q}{2\gamma A_\cH}W_0\left(\frac{128\hbar^3G_4^3}{\gamma^2A_\cH^2\delta^2_q}\right),
\end{align}
where $W_0(x)$ is the Lambert $W$ function. We can see that in the tensionless limit $q\to 1$, 
\begin{align}
    \lim_{q\to1}\delta_q = 0~,
\end{align}
giving the divergent area-law of entanglement entropy \cite{Bombelli:1986rw, Srednicki:1993im} where $\delta_{q}$ can be identified with the UV cutoff appearing in the standard QFT description literature (see \autoref{App:Entropies}). Therefore, the cutoff is related to the geometric orbifold parameter which acquires a physical interpretation in terms of energy scale. 
The finite contribution corresponds to the Gibbons--Hawking entropy in agreement with \cite{Nomura:2017fyh, Dong:2018cuv, Arias:conf}. 
One can also compare the result with the one-loop contributions to black-hole entropy obtained in \cite{Solodukhin:1994yz} by considering conical singularities associated to time coordinate. These conical singularities modifies the temperature of the black hole giving quantum corrections to the entropy which contains two divergences; one associated to interactions near the tip of the cone, and other related to UV divergences. 
In order to obtain the correct result one needs to match the regulators for both divergences.
In particular, for the presented case, we have only one regulator associated to both, the UV divergences and the conical defects. 

\subsection{Dimensionality formulae for large and small $q$}

The limits $q\to\infty$ and $q\to0$ of the Renyi entropy, referred to, respectively, as the min-entropy and the max-, or Hartley, entropy, yield information of the dimensionality of the density matrix $\rho_A$ \cite{Headrick:2010zt,Hung:2011nu}, namely
\begin{align}
    S_{\infty} ={}& -\log\lambda_1~,\\ S_0 ={}& \log \cD~,
\end{align}
where $\lambda_1$ is the largest eigenvalue that is active in $\rho_A$, and $\cD$ corresponds to the number of non-vanishing elements of $\rho_A$. 
In terms of \eqref{min} and \eqref{zero} we get
\begin{align}
    \lambda_1 ={}& (\gamma S_{\rm GH})^{\tfrac32}\exp\{-\gamma S_{\rm GH}\}~, \\ \cD ={}& \exp\{\gamma S_{\rm GH}\}~,
\end{align}
because of the large value of $S_{\rm GH}$  (for example, see \cite{Podolskiy:2018fwk}) we get $\lambda_1 \ll 1$ which is consistent with the fact that $S_\infty$ represents the minimum value of the Renyi entropy,  and $\cD \gg 1$ gives the dimensionality of Hilbert space associated to a single Rindler observer $\cH_A$ such that it has dimension $\exp\{\gamma S\}$, as is already proposed in \cite{Banks:2000fe, Witten:2001kn} when $\gamma = 1$. 
The limit $q\to 0$ corresponds to the limit in which the cutoff moves to the infrared $\delta\to\infty$ leading to the properties of dS spacetime in the IR.

\section{Codimension-2 holography in large $q$ limit}\label{Sec:cod2}

In \cite{Arias:liouv}, it was observed that in the limit $q\rightarrow \infty$, the orbifold dS$_{q}$ reduces to global, three-dimensional de Sitter spacetime (dS$_3$) and the defects are sent to ${\cal{I}}^\pm_3$. 
Moreover, it was shown that Liouville theory on $\Sigma_\xi$ was a consistent truncation at the classical level, and the associated central charge reproduces the one of \eqref{B} for $\gamma=1/2$. Upon dimensional reduction of Newton's constant, the sum of the central charges of the two Rindler observers resembles the one appearing in the context of dS/CFT correspondence \cite{Lin:1999gf, Banados:1998tb, Strominger:2001pn}.
As a matter of fact, this can be viewed as a realization of dS$_3$/CFT$_2$ holography starting from the higher-dimensional singular spacetime dS$_4/\mathbb{Z}_q$. Thus, we will extract information of the two-dimensional CFTs living at the space-like boundaries of three-dimensional de Sitter space $\cI^\pm$. 
In particular, the quantum corrections to the corresponding two-dimensional holographic CFT are thus given by the large-$q$ limits of those computed in the previous section, which remain finite, and have been computed in \cite{Carlip:2000nv}. 
The large-$q$ limit of the embedding coordinates \eqref{embedding1} is given by
\begin{align}
Y_0 = \sqrt{\ell^2-r^2}\cos\theta\sinh(t/\ell)~,\qquad Y_1 = \sqrt{\ell^2-r^2}\cos\theta\cosh(t/\ell)~, \nonumber \end{align}\begin{align}
    Y_2 = r\cos\theta~,\qquad Y_3 = \ell\sin\theta~,\qquad  Y_4 = 0~,
\end{align}
upon using the $\mathbb{Z}_q$ action, satisfying the hyperboloid constraint $Y_A Y^A = \ell^2$ for $A=0,\dots,3$~, resulting in the induced line element
\begin{align}
    ds^2 = \cos^2\theta\left(-f(r)dt^2 + \frac{dr^2}{f(r)}\right) + \ell^2 d\theta^2~.
\end{align}
As proposed in \cite{Arias:entanglement}, the double Wick rotation $t\to i\ell\hat{\varphi}~, \theta\to i\hat{\tau}/\ell$ resembles dS$_3$ in global coordinates, which in our case the embbeding coordinates lead to the line element
\begin{align}
    ds^2 = -d\hat{\tau}^2 +\cosh^2(\hat{\tau}/\ell)\left(\frac{dr^2}{f(r)} + \ell^2 f(r)d\hat{\varphi}^2\right)~,\qquad \hat{\tau}\in\mathbb{R}~,\qquad \hat{\varphi}\in[0,2\pi]~,
\end{align}
where the line element at fixed $\hat{\tau}$ corresponds to a Euclidean $\Rindler_2$, that is, a round $S^2$. 
Thus, taking $r=\ell\cos {\hat{\phi}}$, we get 
\begin{align}
    ds^2 = -d\hat{\tau}^2 + \ell^2\cosh^2(\hat{\tau}/\ell)d\hat{\Omega}_2^2~,\qquad \hat{\phi}\in [0,\pi]~,
\end{align}
where $\d\hat{\Omega}_2^2 = d\hat{\phi}^2 + \sin^2\hat{\phi}~d\hat{\varphi}^2$ is the line element of the unit two-sphere. 
As expected, in the $q\to\infty$ limit, the minimal surfaces $\Sigma_\xi$ corresponds to the past and future conformal infinities ${\cal I}^\pm_3$ of dS$_3$.
The large $q$ limit is illustrated in Figure~\hyperlink{Fig:3}3. 
\vspace{0.7cm}
\begin{center}
\begin{tikzpicture}[scale=0.7]
\draw [thick] (-4,0) circle (2cm);
\node at (-5.5,2) {\footnotesize$S^2$};
\draw[thick, dashed] (-2,0) 
arc[start angle=0,end angle=180, 
x radius=2, y radius=0.5];
\draw[thick] (-2,0) 
arc[start angle=0,end angle=-180, 
x radius=2, y radius=0.5];
{\color{darkblue}
\node at (-0.5,0.45) {\footnotesize{$S^2/\mathbb Z_q$}};
\draw [thick, ->] (-1.3,0) -- (0.35,0);
}
\fill[fill=yellow!10] (2,2.3)--(1,1.7)--(6,1.7)--(7,2.3);
\fill[fill=blue!10] (2,-1.7)--(1,-2.3)--(6,-2.3)--(7,-1.7);
\draw [thick] (2,2.3) --(7,2.3);
\draw [thick] (1,1.7) --(6,1.7);
\draw [thick] (1,1.7) --(2,2.3);
\draw [thick] (7,2.3) --(6,1.7);
\coordinate (N) at (4,2);\coordinate (S) at (4,-2);
\node [darkred] at (N) {\textbullet};
\node [darkred] at (S) {\textbullet};
\draw[thick] (N)to[out=-20,in=20](S);
\draw[thick] (N)to[out=-150,in=150](S);
\draw[thick, dashed] (5.1,0) 
arc[start angle=0,end angle=180, 
x radius=1.05, y radius=0.3];
\draw[thick] (5.1,0) 
arc[start angle=0,end angle=-180, 
x radius=1.05, y radius=0.3];
\draw [thick] (2,-1.7) --(7,-1.7);
\draw [thick] (1,-2.3) --(6,-2.3);
\draw [thick] (1,-2.3) --(2,-1.7);
\draw [thick] (7,-1.7) --(6,-2.3);
\draw [thick, ->] (4.05,0) -- (5.05,0);
\node at (6,0) {\footnotesize{$\ell_q=\frac{\ell}{q}$}};
\node at (7,1.7) {\footnotesize $\Sigma_N$};
\node at (1,-1.7) {\footnotesize $\Sigma_S$};
{\color{darkblue}
\node at (8.4,0.35) {\footnotesize{$q\to\infty$}};
\draw [thick, ->] (7.5,0) -- (9.3,0);
}
\fill[fill=yellow!10] (10,2.3)--(9,1.7)--(14,1.7)--(15,2.3);
\draw [thick] (10,2.3) --(15,2.3);
\draw [thick] (9,1.7) --(14,1.7);
\draw [thick] (9,1.7) --(10,2.3);
\draw [thick] (15,2.3) --(14,1.7);
\fill[fill=blue!10] (10,-1.7)--(9,-2.3)--(14,-2.3)--(15,-1.7);
\draw [thick] (10,-1.7) --(15,-1.7);
\draw [thick] (9,-2.3) --(14,-2.3);
\draw [thick] (9,-2.3) --(10,-1.7);
\draw [thick] (15,-1.7) --(14,-2.3);
\draw [thick, dashed] (12,-2) --(12,2);
\node at (14,0) {\footnotesize{dS$_3$}};
\draw [thick, ->] (12,2) -- (12,1);
\node at (11.7,1.3) {\footnotesize{$z$}};
\node[text width=13cm, text justified] at (5,-5) 
{\small {\hypertarget{Fig:3}\bf Fig.~3}: 
\sffamily{
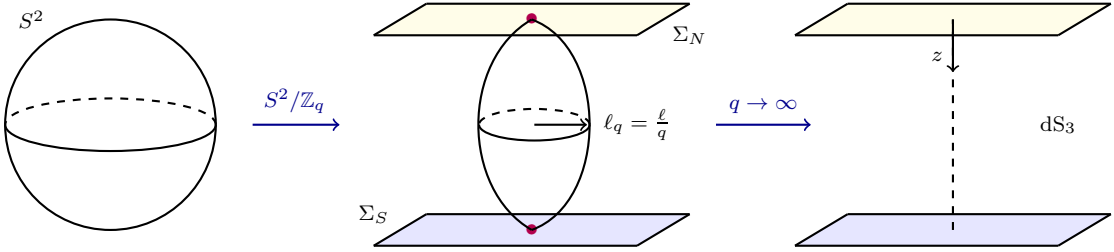
The $q\to\infty$ limit of the geometry corresponds to the zero radius $\ell_q$ limit where the spindle shrinks to a single traverse dimension between the defects. The resulting geometry corresponds to global dS$_3$ spacetime. 
}};
\end{tikzpicture}
\end{center}
\vspace{0.5cm}

As in \cite{Arias:entanglement}, the central charge \eqref{c} of the defects reproduces the result of Strominger \cite{Strominger:2001pn} for dS$_3$, which is
\begin{align}
    c({\Sigma_N \cup \Sigma_S}) \stackrel{\substack{q\to\infty\vspace{-1mm}\\{}}}{\longrightarrow} c({\cal I}^+_3 \cup {\cal I}^-_3) =  \frac{3\ell}{2\hbar G_3}~,
\end{align}
by using the relation $G_4 = 4\ell G_3$~, and was shown to reproduce the three-dimensional de Sitter entropy by taking the large $q$ limit of the modular free energy of the total space given by the Cardy entropy of both observers,
\begin{align}
    S_{\rm C}({\Rindler_N\cup\Rindler_S}) \stackrel{\substack{q\to\infty\vspace{-1mm}\\{}}}{\longrightarrow}  \frac{\pi\ell}{2\hbar G_3}~.
\end{align}

Considering now the quantum corrections of the Cardy entropy on the defects, obtained in \autoref{Sec:Log}, in the large $q$ limit we get
\begin{align}
     S_{\rm C}({\Rindler_N\cup\Rindler_S}) \stackrel{\substack{q\to\infty\vspace{-1mm}\\{}}}{\longrightarrow}  \frac{\pi\ell}{2\hbar G_3} - 3\log \frac{\pi\ell}{2\hbar G_3}+{\rm const}~,
\end{align}
giving the recently obtained first quantum correction of three-dimensional de Sitter \cite{Anninos:2020hfj, Coleman:2021nor}. 
In this limit the dual theory becomes non-thermal, as $T = q^{-1}\to 0$~, in agreement with the interpretation of \cite{Klemm:2001ea, Klemm:2002ir, Arias:entanglement} as dS$_3$ entropy being the Liouville momentum avoiding the problems pointed out in \cite{Dyson:2002nt}.
A natural generalization of this construction would be the inclusion of higher co-dimensional defects. This has been explored in Anti-de Sitter space by considering a causal wedge of AdS with defects generalizing the idea of AdS/CFT \cite{Akal:2020wfl,Miao:2020oey, Geng:2020fxl,Miao:2021ual}.

\section{Conclusions and outlook}

In this work, following the ideas of \cite{Arias:entanglement, Arias:liouv}, we consider Einstein gravity on the quotient dS$/\mathbb{Z}_q$ with Nambu--Goto improvement terms at defects so as to obey the variational principle \cite{Fursaev:1995ef}, such that the smooth de Sitter geometry re-appears in the tensionless limit $q\to 1$.
We find that the cosmological horizon is stabilized by the asymptotic Killing vector fields of \cite{Carlip:1999cy, Silva:2002jq}, inducing an asymptotic symmetry algebra \`a la Barnich--Brandt \cite{Barnich:2001jy} given by a single copy of the Virasoro algebra with non-trivial central charge. We fix the $q$-dependence of the central charge by requiring maxima disorder, resulting in a $q$-dependent Cardy entropy. 
We also choose the remaining free parameter, which was present already in the construction \cite{Carlip:1999cy}, by demanding constant Renyi entropy, which reproduces the semi-classical result \cite{Arias:entanglement, Arias:liouv}, leading to the conclusion that the Gibbons--Hawking entropy can be interpreted as the entanglement entropy between causally disconnected Rindler patches. 
Following \cite{Carlip:2000nv}, we consider Gaussian fluctuations in the density of states in the thermal CFT describing the horizon, which results into finite logarithmic corrections for $q>1$~. 
In the $q\to1$ limit of the Renyi entropy, namely, the tensionless limit, we obtain a divergent von Neumann entropy satisfying the entanglement area law of QFTs \cite{Srednicki:1993im}. We regulate the UV divergence by introducing a cutoff given as a function of the orbifold parameter $q$, allowing us to reinterpret this geometric construction in terms of the energy scale of the theory. 
We also study the $q\to 0$ limit, in which the cutoff moves into the IR, that gives the logarithm of the dimension of the associated Hilbert space, and the $q\to \infty$ limit which gives the larges eigenvalue of the reduced density matrix.
The former indeed reproduces Bank's proposal \cite{Banks:2000fe} that a quantum theory of gravity in an asymptotically de Sitter spacetime has a Hilbert space of finite dimension that equals the logarithm of the Gibbons--Hawking entropy. 

We have studied quantum logarithmic corrections of the Cardy entropy of the Liouville field theory that was obtained in \cite{Arias:liouv} by considering a reduction of the improved Einstein action to the defects. 
In the large $q$ limit, the central charges of this model reproduce the dS$_3$/CFT$_2$ results of \cite{Strominger:2001pn}, in which the central charges of the defects now correspond to those of the conformal asymptopia, and the corresponding Cardy entropy contains logarithmic corrections and reproduces those of three-dimensional de Sitter quantum gravity found by computing the one-loop correction to its partition function on the round three-sphere \cite{Anninos:2020hfj}. 

It would be interesting to match the results of the current work with those found in \cite{Aros:2010jb}, where it was found that the logarithmic correction to entropy is a Noether charge given in terms of three ingredients: the horizon Euler characteristic, the a-trace anomaly appearing the quantum corrections of Einstein field equations, and the solution $\phi_\cH$ to the uniformization problem for $Q$-curvature \cite{malchiodi2007conformal}.
Particularly, these corrections were computed for Schwarzschild and massless topological AdS black holes.
In the current work, we show that the latter corresponds to the double Wick rotation of the static patch in dS spacetime. 
Therefore, the solution for $\phi_\cH$ is the same as in the case of massless, topological black holes, while the Euler characteristic has opposite sign.
Since both are conformally flat backgrounds, it is possible to obtain information about the trace anomaly on the four-dimensional dS background using this formalism and comparing with our result. 

Finally, similar spacetimes have been recently studied in supergravity with negative cosmological constant. 
In particular, the resulting near-horizon geometry of branes wrapping spindles corresponds to a direct product of the type $AdS \times \Sigma$ where $\Sigma$ is a spindle \cite{Ferrero:2020laf,Ferrero:2021wvk,Faedo:2021nub}. 
For instance, it is argued that the near-horizon limit of a M2-brane wrapped around a spindle is described by a well-known classical solution, the Plebanski-Demianski metric \cite{Ferrero:2020twa,Cassani:2021dwa}. 
Here, the tension of the cosmic string produces the acceleration of the black hole \cite{Appels:2017xoe,Gregory:2017ogk} which interplays within the cosmological constant creating a Rindler horizon if the acceleration is large enough. 
This second horizon resembles the characteristics of the cosmological horizon as it disconnects the regions of the spacetime preventing access to the AdS boundary. 
The thermodynamics of these spacetimes seems to be well-understood in the regime where the acceleration is \textit{slow} and the Rindler horizon vanishes \cite{Appels:2017xoe,Anabalon:2018ydc,Gregory:2019dtq,Arenas-Henriquez:2022www}.
Nevertheless, a consistent thermodynamic description of the accelerating horizon is still unclear, and one could expect that using the asymptotic symmetry algebra of this horizon as in \autoref{Sec:CentralCharge} could lead to a better understanding of accelerating black holes and their holographic description.
We leave this question open for future work. 

\paragraph{Acknowledgements.} We would like to thank C. Arias, R. Aros, A. Faraggi, C. Far\'ias, C. Iazeolla, K. Lara, N. Nguyen, M. Valenzuela, and B. Vallilo for enlightening discussions on various topics related to this work. We also want to thank to K. Narayan for his interesting comments regarding the notion of entanglement in de Sitter. 
{\sc Ps} acknowledges the kind support of the Centro de Ciencias Exactas at Universidad del Bio-Bio. 
The work of {\sc Gah} is funded by {\sc Becas Chile} (ANID) Scholarship N$^{\rm o}~ 72200271$. 

\begin{appendix}

\section{Entropy functions}\label{App:Entropies}

A key distinguishing feature of quantum mechanics vis-\`a-vis  classical mechanics is the notion of quantum entanglement, whereby a system built from localizable degrees of freedom attached to two separate regions, $A$ and $B$, say, of a manifold, or a set of points, is assigned a direct-product Hilbert space $\mathfrak{H}_A\otimes \mathfrak{H}_B$, which contains \emph{pure} quantum states
\footnote{In classical mechanics, the system is assigned a direct-product symplectic manifold $M_A\times M_B$; probability distributions on this space can describe mixed classical states in which the values observables in $A$ and $B$ are correlated, but there is no classical counterpart of quantum entanglement at the level of pure, classical states, which are given by delta functions; indeed, pure, classical states remain pure as one of the symplectic manifolds are integrated out. 
Moreover, defining a pure classical state as a delta-sequence, its von Neumann entropy tends to minus infinity as the volume of support goes to zero, manifesting a classical ``catastrophe''.} given by superpositions that correlate observables confined to $A$ and $B$, that is, the eigenvalues of operators of the form ${\cal O}_A\otimes {\rm Id}_B$ and ${\rm Id}_A\otimes {\cal O}_B$.
Thus, an observer in one of the two regions, $B$, say, measuring ${\cal O}_B$ by performing a von Neumann measurement, which eventually produces a pure state in $\mathfrak{H}_A\otimes \mathfrak{H}_B$ with fixed eigenvalue of ${\rm Id}_A\otimes {\cal O}_B$, bears an impact on the spectrum of ${\cal O}_A\otimes {\rm Id}_B$ in region $A$ (independently of the scale of the spatial separation).

\paragraph{Renyi entropy and modular free energy.} Consider the reduced density matrix
\be
\rho_A := {\rm Tr}_{\mathfrak{H}_B} \,\rho_{AB}~,
\ee
whose spectral properties manifest the degree of entanglement of the original pure state.
These properties are encoded into the spectral function
\begin{align}
    S_{q,A} := \frac{1}{1-q}\log \cZ_A\ ,\qquad \cZ_A:= {\rm Tr}_{\mathfrak{H}_A}\,(\rho_A)^q~,\qquad {\rm Re}(q)>0\ ,
\end{align}
known as the Renyi entropy; when restricted to the strictly positive integers, the parameter $q$ is referred to as the replica number.
The Renyi entropy is $q$-independent iff the reduced density matrix describes an equiprobable ensemble, in which case the original state is said to be maximally entangled (with respect to the separation of the point set into $A\cup B$).
The limits
\begin{align}
    S_{{\rm E},A} := \lim_{q\to1}{S}_{q,A}\ ,\qquad  S_{0,A}:=\lim_{\q\to 0} {S}_{q,A}
     \ ,\qquad S_{\infty,A} := \lim_{q\to\infty} {S}_{q,A}\ ,
\end{align}
are referred to, respectively, as the entanglement, Hartley and minimum entropies of $\rho_A$.
From 
\be
    S_{{\rm E},A} = -{\rm Tr}_{\mathfrak{H}_A}\, \rho_A\,\log \rho_A\ ,\ee
it follows that the entanglement entropy is the von Neumann entropy of $\rho_A$, which serves as a good measure of the degree of entanglement in the original pure state\footnote{Viewing the reduction operations as processes taking place in a modified, nonlinear version of quantum mechanics, one may think of quantum dynamics acts as the fundamental source of (entanglement) entropy.}.
The Hartley entropy can be identified as 
\be  S_{0,A}:= \log\cD_A~,\ee
where $\cD_A$ is the number of non-zero eigenvalues of $\rho_A$, that is, the dimension of the subspace of $\mathfrak{H}_A$ that participates in the entanglement.
Assuming that the reduced density matrix is in the image of the exponential map, viz. 
\be
\rho_A =: \exp\left\{-H_A\right\}~,
\ee
where $H_A$ is referred to as the modular Hamiltonian, one has
\be S_{\infty,A} := \lim_{q\to\infty} {S}_{q,A} \equiv -\log\lambda^{\rm max}_A~,\ee
where $\lambda^{\rm max}_A$ is the largest eigenvalue of $\mathfrak{H}_A$.
The existence of such a Hamiltonian implies that the Renyi entropy can be identified as
\be S_{q,A}=\left(1-\frac{1}q\right)^{-1} F_A\ ,\qquad F_{A} := - \frac{1}q \log\cZ_{A}\ ,\label{BaezApp}\ee
where $F_A$ is thus the free energy of the corresponding thermal system with temperature, entropy, and thermal energy given by
\begin{align}\label{modS}
     T_q:= q^{-1}\ ,\qquad {S}_A := (1-q\partial_q)\log\cZ_{A}\ ,\qquad E_A= -\partial_q\log\cZ_A~\ ,
\end{align}
respectively (and one has $\lim_{q\to1} {S}_A=S_{{\rm E},A}$); this system is often referred to as the modular system, and $F_A$, $T_q$, $S_A$ and $E_A$ as the modular free energy, temperature, modular entropy and modular energy, respectively.
The Renyi and modular entropies are related by
\begin{align}
    {S}_A= q^2 \partial_q\left(\frac{q-1}{q}S_{q,A}\right)~.
\end{align}
Inverting this relation yields 
\begin{align}
    S_{q,A} = \frac{q}{q-1}\int_1^q \frac{d\tilde{q}}{\tilde{q}^2}{S}_A~.
\end{align}

\paragraph{Cardy--Calabrese formula and orbifolds.} In quantum field theories on metric backgrounds, the modular partition function can be computed by letting $\rho_A\equiv \exp\{-H_A\}$ represent the foliation of a smooth, Euclidean geometry $(M_A,ds^2)$, with $M_A\supset A$, generated by $H_A$ in one unit of Euclidean time, with $A$ treated as a co-dimension one Cauchy surface; this yields the Cardy--Calabrese formula
\be
    \cZ_A= \frac{{\cZ}_A\left[\widehat{M}_{A,q},d\hat s^2\right]}{\left(\cZ_A\left[{M}_{A},ds^2\right]\right)^q}~,
\ee
where ${\cZ}_A\left[\widehat{M}_{A,q},ds^2\right]$ is computed on the $q$-fold, branched, Euclidean cover geometry of $(M_A,ds^2)$, with $(\widehat{M}_{A,1},d\hat s^2)\equiv (M_A,ds^2)$, constructed by gluing together $q$ copies of $(M_A,ds^2)$ along $A$, which yields a branched geometry with a smooth metric $d\hat s^2$, and a natural $\mathbb{Z}_q$-action with ramification surface given by the entangling surface $\partial A$.
Thus, the Renyi entropy
\be
    S_{q,A} = \frac{1}{1-q}\left(\log {\cZ}_A\left[\widehat{M}_{A,q},d\hat s^2\right]-q\log \cZ_A\left[{M}_{A},ds^2\right]\right)~.
\ee
In the semi-classical limit, one has
\be {\cZ}_A\left[\widehat{M}_{A,q},d\hat s^2\right]\approx \left.\exp\left\{-\frac1{\hbar} I_{\rm E}\left[\widehat{M}_{A,q},d\hat s^2\right]\right\}\right|_{\rm Saddle}\ ,\ee
where $I_{\rm E}$ is the Euclidean action and the saddle-point requires a suitable boundary condition; assuming locality, it follows that 
\be I_{\rm E}\left[\widehat{M}_{A,q},d\hat s^2\right]= q  I_{\rm E}\left[(\widehat{M}_{A,q},d\hat s^2)/\mathbb{Z}_q\right]\ ,\ee
where the orbifold geometry 
\be (\widehat{M}_{A,q},d\hat s^2)/\mathbb{Z}_q=(M_A,ds^2_q)\ ,\ee
that is, it is given by the manifold $M_A$ equipped with a non-smooth metric $ds^2_q$ with a conical singularity of co-dimension two at $\partial A$ with a deficit angle $2\pi(1-q^{-1})$. 
The quantum corrections to Renyi entropy contains short-distance divergences inside the bulk, that are independent of entanglement data, and at the border of the traced-out region, referred to as the entanglement surface; the latter exhibit universal scaling behaviours leading to area laws for the entanglement entropy, viz.
\begin{align}
    S_{{\rm E}, A} = \gamma\frac{A(\partial A)}{\delta^{d-2}}(1+o(\delta))\ ,\qquad \gamma \in \mathbb{R}^+~,
\end{align}
where $\delta$ is a UV cut-off, and $\gamma$ depends on the field content of the theory. 

At high temperature, the highest UV-finite contribution is dominated by the thermal entropy of the subset $A$ \cite{Cardy:2014jwa}, and consdering $B$ to be empty, the entanglement entropy corresponds to the thermal entropy
\begin{align}
    S_{{\rm E}, A} = \beta\langle H_A\rangle + \log\cZ_A = \beta(E_A - F_A) = S_{{\rm th},A}~.
\end{align} 

\paragraph{$q$-derivatives.} Defining  
\begin{align}
    \left(\frac{\partial F}{\partial x}\right)_{q} \equiv \frac{F(qx) - F(x)}{qx-x}~,
\end{align}
referred to as $q$-derivative of $F(x)$, which recovers the standard derivative wehen $q=1$~, the Renyi entropy can be seen as the $q^{-1}$-derivative of the negative free energy \cite{baez2011renyi}, viz.
\begin{align}
    S_q = -\left(\frac{\partial F}{\partial T}\right)_{q^{-1}}~.
\end{align}
This relation can be generalized by considering a re-scaled temperature 
\begin{align}
    T = \frac{T_0}{q}~,
\end{align}
where we normalize the partition function at $T_0$ as $\widehat{\cZ}(T_0) = 1$~; then
\begin{align}
    S_{q,A} = \frac{T \log\widehat{\cZ}(T) - q \log\widehat{\cZ}(T_0)}{T-T_0}~,
\end{align}
which by normalizing $T_0 = 1$ simplifies to Eq. \eqref{BaezApp} with $F(q) = F_A$~. 
In other words, the Renyi entropy amounts to the maximum work that a system in thermal equilibrium can perform by reducing its temperature by a factor of $q^{-1}$, divided by the change in temperature. 

\paragraph{Disorder and von Neumnann entropy.}
The von Neumann entropy has a natural thermodynamic interpretation as the equilibrium value of a disorder functional.
Let us refer to a set $\{\sigma_\xi\}_{\xi=1}^{N}\equiv \Sigma$ of $N$ labelled elements, or micro states, as an $N$-state system.
Let $\mathbb{R}^N_+$ be the $N$-dimensional real cone.
A map $\rho:\Sigma\to \mathbb{R}^N_+$ obeying 
\be \sum_{\xi=1}^N \rho(\sigma_\xi)=1\ ,\ee
is referred to as a macro-state.
Each disjoint union $\Sigma=\Sigma'_1\cup\Sigma'_2$ induces a 2-state system $\check\Sigma:=\{\Sigma'_1,\Sigma'_2\}$, whose micro states are thus given by the two disjoint subsystems, with a macro state $\check\rho$ defined by
\be \check\rho(\Sigma'_i)=\sum_{\sigma\in \Sigma'_i}\rho(\sigma)\ ,\qquad i=1,2\ .\ee
The disorder functional $D:\rho\to \Real_+$ is defined by requiring that 
\be \label{Disorder}
D\left[\rho\right]= \check\rho(\Sigma'_1)D\left[\frac{1}{\check\rho(\Sigma'_1)}\rho|_{\Sigma'_1}\right]+\check\rho(\Sigma'_2) D\left[\frac{1}{\check\rho(\Sigma'_2)}\rho|_{\Sigma'_2}\right]+ D\left[\check\rho\right]\ .\ee
for any disjoint union, that is, the disorder of $\rho$ is given by the weighted contributions from the two subsystems plus an extra contribution arising due to forming the disjoint union, referred to as the mixing entropy.
The latter is clearly negligible in the large $N$ limit, but significant after repeated partitions down to a large number of subsystems of size of order one, and the sole source of entropy in a process setting out from separate micro states. 
It follows that
\be D[\rho]=k\sum_\xi \rho(\sigma_\xi)\log\frac{1}{\rho(\sigma_\xi)}\ ,\ee
where $k$ is an undetermined constant. 

Classical thermodynamics amounts to maximizing the disorder in the space of macro-states while keeping expectation values of observables fixed; in the absence of any constraints, the extremal macro-state is the equiprobable distribution of micro states, while the presence of constraints imply that the extremal macro-states are given by Boltzmann distributions with conjugate variables arising as Lagrange multipliers.
An interesting question is thus how to modify the above notions such that the resulting extremization procedure leads to Renyi entropy.

\section{Asymptotic charges}\label{App:AQ}
Let $M$ be a manifold with interior $M'$ and boundary $\partial M$ admitting a one-parameter foliation with leafs $\Sigma$ for a canonical formulation, and denote by ${\rm Vect}(M;\partial M)$ the algebra of vector fields on $M$ that stabilizes $\partial M$.
Let i) $g$ be an on-shell metric that equips $\partial M$ with an asymptotic structure $[g]|_{[\partial M]}$; ii) ${\cal C}$ be a space of classical solutions $\hat g:=g+h$ that are smooth in the interior of $M$, containing $\hat g=g$; and iii) $\mathfrak K$ be a Lie algebra represented by subalgebra of ${\rm Vect}(M;\partial M)$, referred to as the AKVs, that stabilizes ${\cal C}$.
The canonical formalism \cite{Barnich:2001jy,Silva:2002jq} assigns the system asymptotic charges
\be \widehat{{\cal Q}}^{(\epsilon)}_{\vec\zeta}[\hat g] =  \oint_{\Sigma_\epsilon}\hat k_{\vec\zeta}[\hat g]\ ,\ee
where $\vec \zeta$ is the vector field representing $\zeta\in \mathfrak K$, and $\Sigma_\epsilon\subset M\stackrel{\epsilon\to0}{\longrightarrow }  \Sigma$, that are diffeomorphism invariant, that is
\be \int_M\left(({\cal L}_{\vec v}\hat g)_{\mu\nu}\frac{\delta}{\delta \hat g_{\mu\nu}}+({\cal L}_{\vec v}\vec\zeta)^\mu \frac{\delta}{\delta (\vec \zeta)^\mu}\right)\widehat{{\cal Q}}^{(\epsilon)}_{\vec \zeta}=0\ ,\ee
for smooth $\vec v$ on $M$, i.e.
\be \int_M ({\cal L}_{\vec v}\hat g)_{\mu\nu}\frac{\delta}{\delta \hat g_{\mu\nu}}\widehat{{\cal Q}}^{(\epsilon)}_{\vec \zeta}=-\widehat{{\cal Q}}^{(\epsilon)}_{[\vec v,\vec \zeta]}\ , \label{B.3}\ee
such that if 
\be \widehat {\cal Q}_{\vec\zeta}:=\lim_{\epsilon\to 0}\widehat{{\cal Q}}^{(\epsilon)}_{\vec \zeta}\ ,\label{B.4}\ee
are finite, then they are moment functionals for a projective representation of $\mathfrak K$ in ${\cal C}$ by means of a Dirac bracket $\{\cdot,\cdot\}$, in which $g$ is treated as a constant background, viz.
\be \{\widehat{\cal Q}_{\vec\zeta}, h\}:={\cal L}_{\vec\zeta} h\ .\ee
It follows that $\{\widehat{{\cal Q}}_{\vec \zeta}, \hat g\}={\cal L}_{\vec\zeta}\hat g- {\cal L}_{\vec\zeta} g$, which yields
\be \{\widehat{{\cal Q}}_{\vec \zeta},\widehat{{\cal Q}}_{\vec \zeta'} \}\equiv \int_M (\widehat{{\cal Q}}_{\vec \zeta})_{\mu\nu}\frac{\delta}{\delta \hat g_{\mu\nu}}\widehat{{\cal Q}}_{\vec \zeta}=\widehat{{\cal Q}}_{[\vec \zeta',\vec \zeta]}-\int_M ({\cal L}_\zeta g])_{\mu\nu}\frac{\delta}{\delta \hat g_{\mu\nu}}\widehat{{\cal Q}}_{\vec \zeta'}\ ,\ee
upon using \eqref{B.3} and when combined with \eqref{B.4}.
Assuming fall-off conditions on $h$ such that
\be \widehat{{\cal Q}}_{\vec \zeta}[\hat g]=\widehat{{\cal Q}}_{\vec \zeta}[g]+{{\cal Q}}_{\vec \zeta}[h;g]\ ,\qquad {{\cal Q}}_{\vec \zeta}[h;g] := \oint_\Sigma k_{\vec \zeta}[h;g]\ ,\ee
where $k_{\vec \zeta}[h;g]$ is linear in $h$, one has
\be  \{{{\cal Q}}_{\vec \zeta},{{\cal Q}}_{\vec \zeta'} \}={{\cal Q}}_{[\vec \zeta',\vec \zeta]}+ C(\vec\zeta,\vec\zeta';g)\ ,\ee
where the central term
\be C(\vec\zeta,\vec\zeta';g)=\widehat{{\cal Q}}_{[\vec \zeta',\vec\zeta]}[g]+\oint_\Sigma k_{\vec \zeta}[{\cal L}_{\vec\zeta'}g;g]\ \ee
which is an anti-symmetric bi-linear form on $\mathfrak K$ built from background charges and the charges of ${\cal L}_{\vec\zeta}g$ treated as fluctuations; as a simple sign check, one has
\be \{{\cal Q}_{\vec \zeta},\underbrace{\{{\cal Q}_{\vec \zeta'},h\}}_{{\cal L}_{\vec\zeta'}h}\}+\{{\cal Q}_{\vec \zeta'},\underbrace{\{h,{\cal Q}_{\vec \zeta} \}}_{-{\cal L}_{\vec\zeta}h}\}+\{h,\underbrace{\{{\cal Q}_{\vec \zeta},{\cal Q}_{\vec \zeta'} \}}_{{\cal Q}_{[\vec \zeta',\vec \zeta]}+ C}\}={\cal L}_{\vec\zeta'}\{{\cal Q}_{\vec \zeta},h\}-{\cal L}_{\vec\zeta}\{{\cal Q}_{\vec \zeta'},h\}-{\cal L}_{[\vec \zeta',\vec \zeta]}h\ ,\ee
i.e. the algebra is indeed consistent with the Jacobi identity.
While the computation of the central charge is based on self-consistency, the existence of the projective representation requires a duality relation \cite{Barnich:2001jy} between the scaling behaviours of elements in ${\cal C}$ and $\mathfrak{K}$ in the asymptotic region, referred to as the asymptotic boundary conditions (ABC).
Relaxing the ABC on ${\cal C}$, enlarges ${\rm Vect}(M;\partial M)_{\cal C}$ and restricts ${\cal C}^\ast_{\widehat{\cal Q}}$, where ${\rm Vect}(M;\partial M)_{\cal C}$ and ${\cal C}^\ast_{\widehat{\cal Q}}$, respectively, denote the stabilizer and ${\cal Q}$-dual of ${\cal C}$ in ${\rm Vect}(M;\partial M)$; thus, imposing ABC such that 
\be {\rm Vect}(M;\partial M)_{\cal C}={\cal C}^\ast_{\widehat{\cal Q}}\ ,\ee
yields symplectic spaces of boundary states with finite central charges containing the orbit of $g$ under the $\mathfrak K$-action.
In a near-horizon geometry, the null Killing vector field $\vec \chi$ (NKV) induces an algebra $\mathfrak K_{\vec \chi}$ of AKVs that preserve the NKV, in its turn inducing ${\cal C}_{\vec\chi}=(\mathfrak K_{\vec \chi})^\ast_{\widehat{\cal Q}}$.

\section{Bunch--Davies vacuum}\label{App:Bunch}
Considering the deformed background \eqref{extended}, we can characterize macroscopic thermal properties of a entangled state that is encoded in the Boltzmann factor $\exp\{-\omega/T_{dS}\}$ where $\omega$ corresponds to the energy eigenstate with temperature $T_{dS}$. Near to the defects, the line element \eqref{extended}
\begin{align}\label{ntd}
    ds^2 \approx -f(r)dt^2 + \frac{dr^2}{f(r)} +\dots~,
\end{align}
and taking the near horizon limit $r\to\ell$, by considering the dimensionless Rindler coordinates $\tau = \ell t$ and $\Upsilon^2 = 2(1-r/\ell)$~,
\begin{align}\label{nh}
    ds^2\approx -\Upsilon^2 d\tau^2 + d\Upsilon^2~, \qquad \Upsilon \ll 1~.
\end{align}
Considering a scalar field \cite{Bunch:1978yq} on the near-to-defect background \eqref{ntd} expanded in eigenstates with energy $\omega$ 
\begin{align}
    \Phi(t,r) = \sum_n \phi_n(r)\exp\{-i\omega t\}~, \qquad \omega >0~,
\end{align}
and taking the near horizon limit, the mode expansion in terms of the Rindler coordinates \eqref{nh} renders
\begin{align}
    \Phi(\tau,\Upsilon) = \sum_n \phi_n(\Upsilon)\exp\{-iN\tau\}~, \qquad N\equiv \ell\omega~,
\end{align}
which turns the Boltzmann factor near the horizon to
\begin{align}
    \exp\left\{-\frac{\omega}{T_{dS}}\right\} = \exp\left\{-\frac{N}{T}\right\}~, 
\end{align}
where thus $N$ corresponds to the number of states with temperature
\begin{align}\label{T}
    T = \frac{1}{2\pi}~,
\end{align}
referred as the Bunch--Davies vacuum or Euclidean vacuum \cite{Bunch:1978yq}. We follow to use the temperature of modes near the horizon \eqref{T} in the Cardy formula.

\end{appendix}

\bibliographystyle{JHEP}
\bibliography{biblio.bib}

\end{document}